# Anomalous, pre-yield grain-boundary sliding in copper revealed with *in-situ* high-resolution strain mapping

Benjamin Poole[1a], David Lunt[1,2], Luke Hewitt[1], Chris Hardie[1],

[1] United Kingdom Atomic Energy Authority, Culham Campus, Abingdon, Oxon, OX14 3DB, UK

[2] Department of Materials, University of Manchester, Manchester, M13 9PL, UK

[a] Corresponding Author: ben.poole@ukaea.uk

**Graphical abstract**

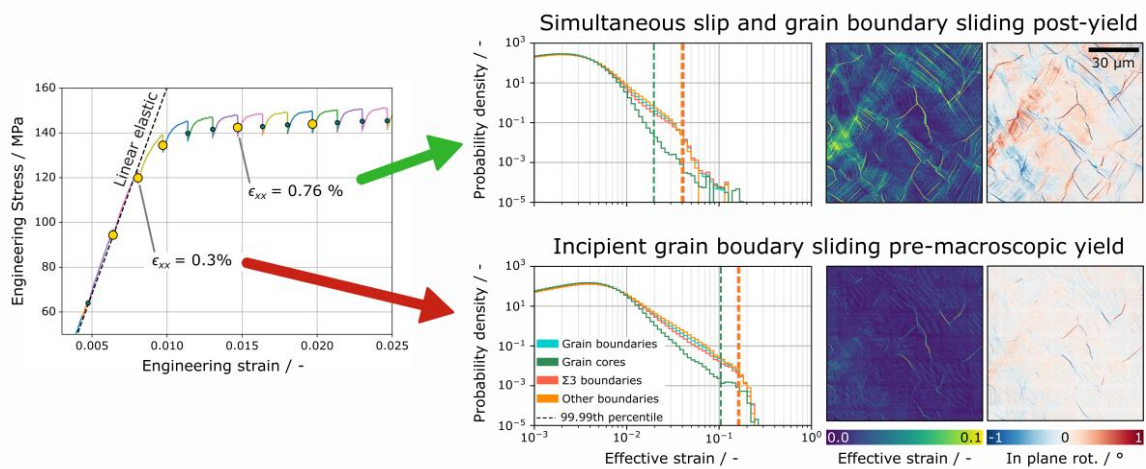

**Abstract**

Grain boundary sliding is typically associated with high temperature deformation in engineering alloys. Here, we examine grain boundary sliding at room temperature in oxygen-free high-conductivity copper under quasi-static tensile testing. By using high-resolution digital image correlation (HRDIC) conducted *in-situ* within a scanning electron microscope to produce time-series strain maps, we unexpectedly observe that grain boundary sliding occurs extensively prior to macroscopic yield, and before the onset of significant crystallographic slip. Extreme values in strain and in-plane rotation are found to be associated with grain boundaries immediately prior to yield and during the initial stages of plastic deformation, which are higher than those associated with crystallographic slip. By combining laser scanning confocal microscopy height mapping with the strain maps and orientation maps from electron backscatter diffraction, grain boundary sliding character is determined, finding evidence of pure in-plane, pure out-of-plane and mixed-mode sliding.

**Main text**

Grain boundary sliding (GBS) is a deformation phenomenon typically associated with elevated temperatures [1–4], in particular above ~0.4 times the homologous temperature [5]. It is often associated with damage during the later stages of creep [6] and, as such, an appreciation of grain boundary sliding during early deformation is likely key to understanding the inherent failure mechanism.

High-resolution digital image correlation (HRDIC) is a technique capable of producing strain maps at the nanoscale resolution that can then be linked to the underlying microstructure to enable deformation mechanisms to be examined at the grain and grain boundary scale. Previous work on copper-base alloys examined with HRDIC has shown evidence of grain boundary sliding at room temperature [7,8], a finding that was somewhat unexpected due to the low experimental temperatures. Grain boundary sliding has also been observed using HRDIC in magnesium [9–11], aluminium [12–14], and, more recently, in thin zinc coatings [15]. Whilst this may be dismissed as simply an artefact of HRDIC only examining the free surface of a specimen, there are numerous HRDIC-based studies where there is little evidence of grain boundary sliding in numerous materials [16]. These are typically high melting point materials including , but not limited to, steels [7,17,18], nickel-base alloys [19,20], zirconium-base alloys [21,22], high entropy alloys [23] and titanium-base alloys [24–26].

Recent advances in performing *in-situ* heating and loading experiments within a scanning electron microscope (SEM) combined with automation [27] have allowed for both increased experimental throughput and, more crucially, the collection of time-series HRDIC datasets. These automated, *in*-situ experiments provide both *where* and *when* strain localisation occurs, revealing which mechanisms are of key prominence at different stages of deformation. However, previous studies observing grain boundary sliding have either been performed *ex-situ,* and therefore could not capture the temporal evolution of the process, or they have not fully exploited the time series data available from *in-situ* testing [3,4,7,8,12,13,28]. We aim to address this knowledge gap by applying automated *in-situ* HRDIC-based testing techniques to pure copper.

For this study, an oxygen-free high-conductivity copper tensile specimen fabricated using wire electric discharge machining was used. Metallographic surface preparation was conducted as described in [8], producing a deformation free surface suitable for both electron backscatter diffraction (EBSD) measurements and high-resolution digital image correlation patterning. The speckle pattern was then applied to the surface using the procedure outlined in [7]. Briefly, a 3 nm layer of titanium, followed by a 5 nm layer of silver was applied to the surface and then remodelled for 120 minutes with a 1 wt.% NaBr in isopropanol solution. Here, we exploited a key property of this pattern, namely the ability to perform EBSD through the patterned surface [7]. By applying the pattern *before* EBSD measurement, the experimental procedure was simplified.

All scanning electron microscopy was performed using a TESCAN CLARA field emission (FE) gun SEM equipped with and Oxford Instruments Symmetry S3 EBSD detector. A schematic diagram of the entire experimental workflow is detailed in Figure 1. Following speckle patterning, the sample was loaded into the SEM. A region of interest (ROI) was selected and marked using a platinum fiducial marker, in the form of a cross, deposited using electron beam deposition with the instrument's gas injection system. EBSD orientation measurements (20 kV accelerating voltage, 10 nA probe current, 200 nm step size) were recorded over a 750 × 750 µm region to capture an ROI containing several thousand grains.

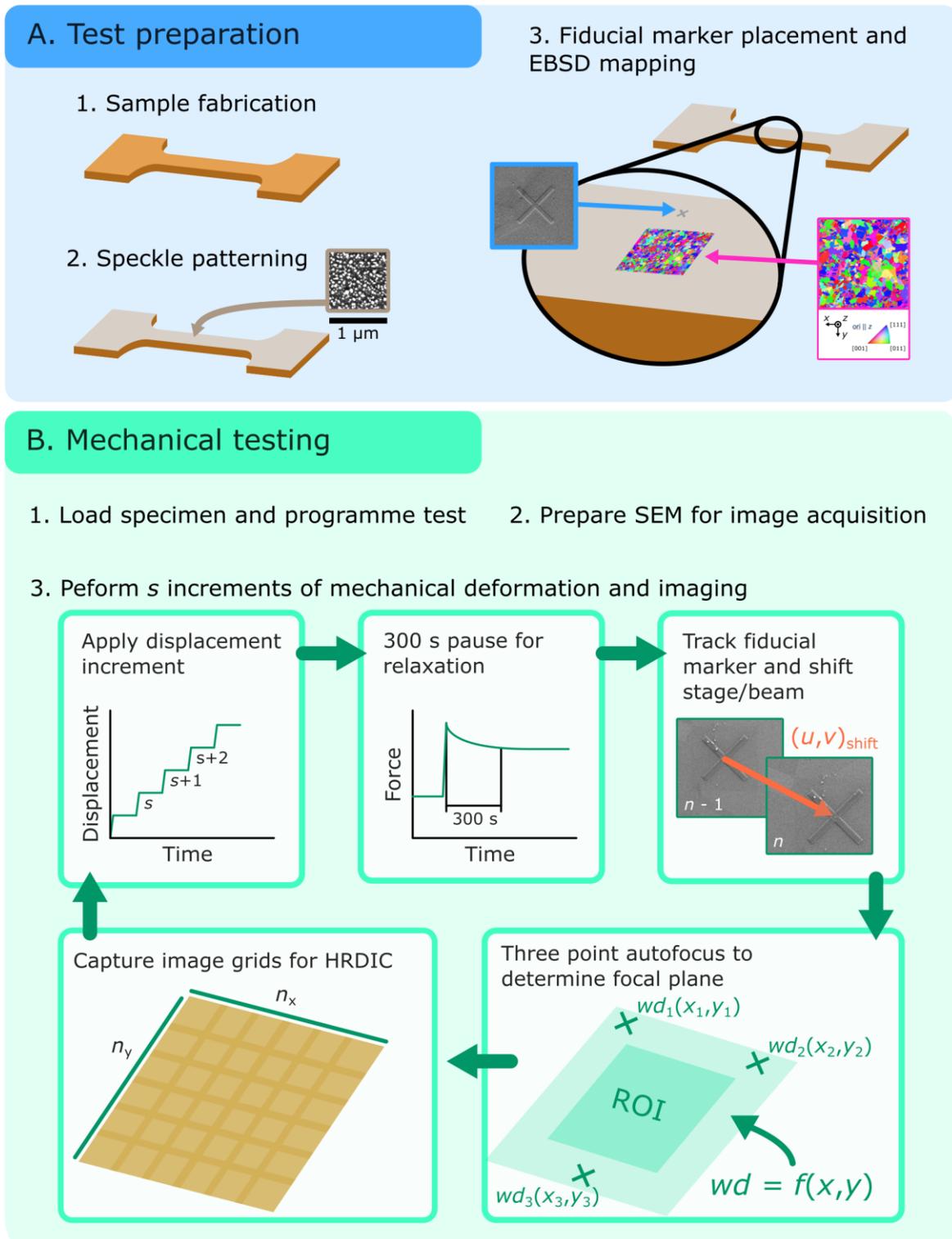

Figure 1: Experimental workflow with both sample preparation and mechanical testing phases.

In-situ mechanical loading was then performed using a NewTec Scientific MT1000 microtester within the same TESCAN CLARA FESEM. During testing, the microtester control software *Softstrain* takes control of both the mechanical loading sequence through the MT1000 microtester and all electron microscope functionality through the TESCAN Essence application programming interface. As such, fully automated testing is possible, facilitating both the high spatial and temporal resolution key to the current experiment that would not be possible through manual instrument control [27].

For each deformation step, a 19 × 19 grid of 20 μm × 20 μm tiles was captured. Image grids were then fused into a single composite image using ITK-montage [29]. Digital image correlation was subsequently performed by correlating all deformation steps relative to the step before deformation following the test using pyvale [30], with a reducing window size fast Fourier transform-based algorithm. The settings and parameters for the digital image correlation process are given in Table 1. Following testing, HRDIC and EBSD datasets were linked using DefDAP [31].

**Table 1: DIC processing parameters**

| Parameter | Value |
| --- | --- |
| Software | pyvale v2026.1.3 |
| Subset size | 41 px × 41 px |
| Subset step | 10 px |
| Correlation algorithm | 2D DIC, Fast Fourier Transform, reducing window size |
| Tile field of view | 20 μm × 20 μm / 2048 px × 2048 px |
| Montage field of view | 300 μm × 300 μm |
| Capture instrument | TESCAN CLARA FESEM, ET-SE detector |
| Montaging software | ITK-Montage 0.8.2, image linear blending |
| Strain field resolution | 98 nm |
| Differentiation method | Second order accurate central differences |

Maps of effective strain, defined as $E_{\text{eff}} = \sqrt{\left(\frac{F_{11}-F_{22}}{2}\right)^2 + \left(\frac{F_{12}+F_{21}}{2}\right)^2}$, and in-plane rotation under the maximum applied displacement are presented in Figure 2. Overall deformation trends are broadly concordant with those previously observed in copper, with regions of diffuse slip, regions of highly localised slip and evidence of grain boundary sliding [7,8,32]. Trans-granular bands of deformation are visible across the map at 45° to the horizontal loading direction, manifesting as channels of intense, finely spaced slip bands with a common in-plane rotation direction. Grain boundary sliding is observable as bands of elevated effective strain and highly localised rotation.

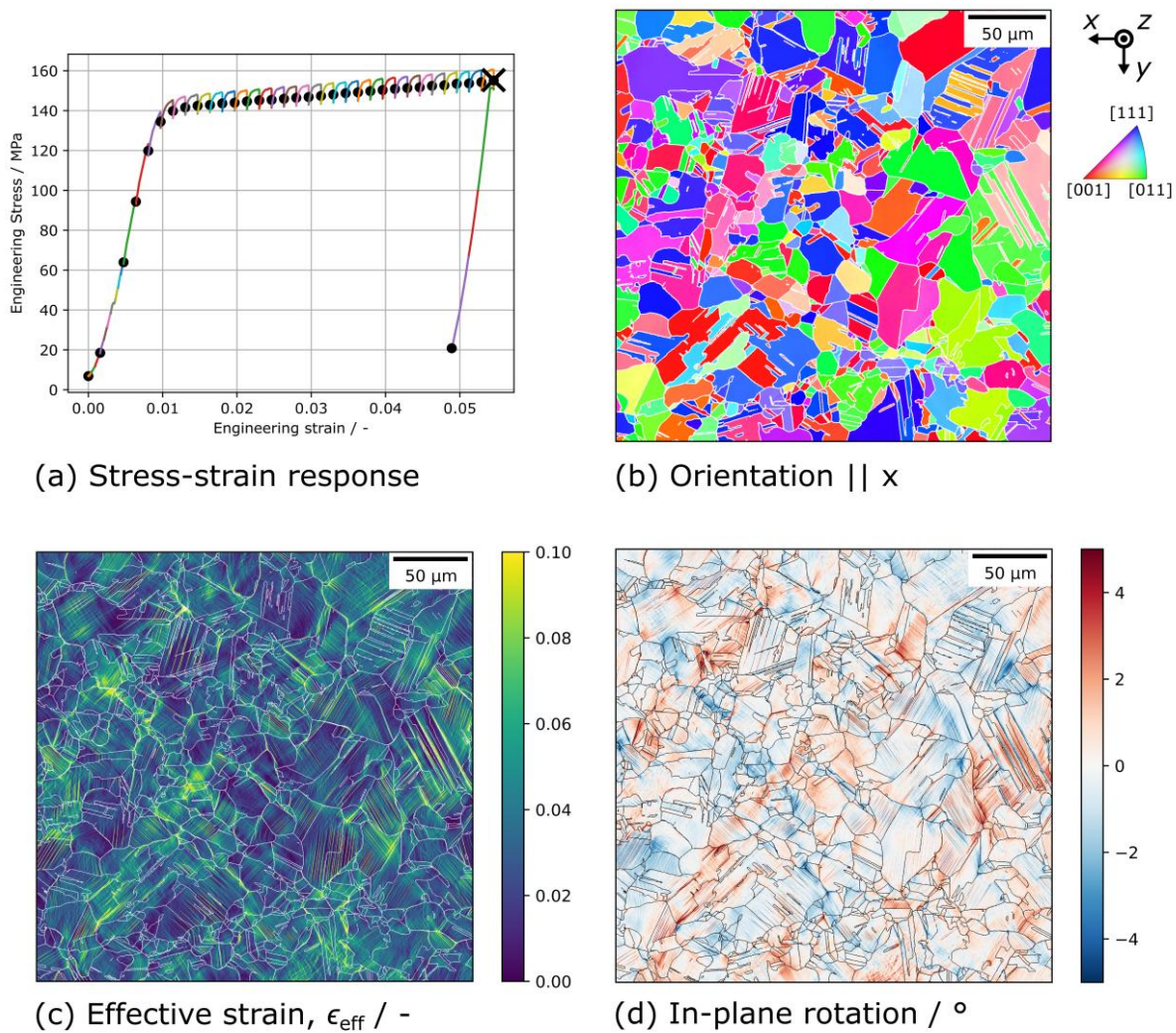

**Figure 2: In-situ experimental output.** The engineering stress-strain curve (a) shows individual automated testing steps in different colours with image montage points shown as black circles. The orientation map (b) shows orientations parallel with the primary loading direction (*x*-direction). (c) and (d) are the HRDIC derived effective strain and in-plane rotation maps, respectively, taken under the maximum load condition indicated with an × symbol in (a).

The key focus of this study is the onset of deformation during the nominally elastic regime and close to the yield point. We define the macroscopic yield point as the point at which the stress-strain response deviated from linearity. Five deformation snapshots are shown in Figure 3, captured at the following points: approximately 70% of the yield stress, at the macroscopic yield point, immediately following yield, and two points post-yield. The first occurrences of grain boundary sliding are visible well before macroscopic yield (Figure 3c, $\epsilon_{11}$ = 0.21%) with minimal deformation activity away from grain boundaries. This trend continues until the macroscopic yield point (Figure 3c, $\epsilon_{11}$ = 0.3%) with clear grain boundary deformation but only faint, diffuse deformation within the grain cores. Following yield, and as expected, slip becomes prevalent, initially as diffuse plasticity, followed by the nucleation and growth of more intense slip bands. However, grain boundary sliding intensity, as measured through strain and rotation intensity continues to increase monotonically with increasing applied strain.

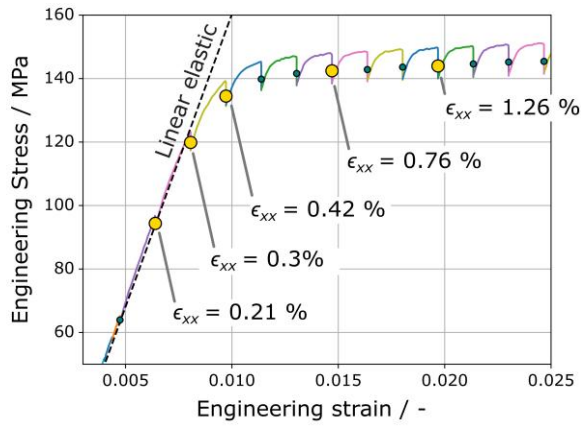
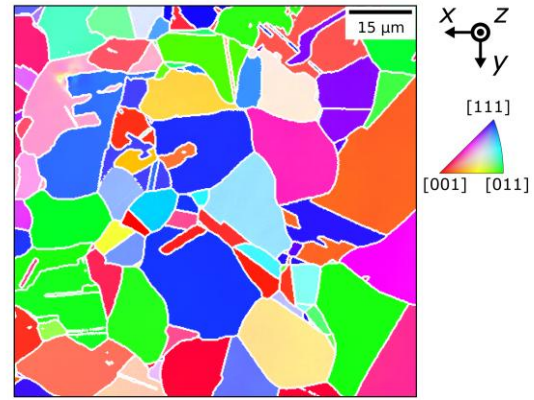
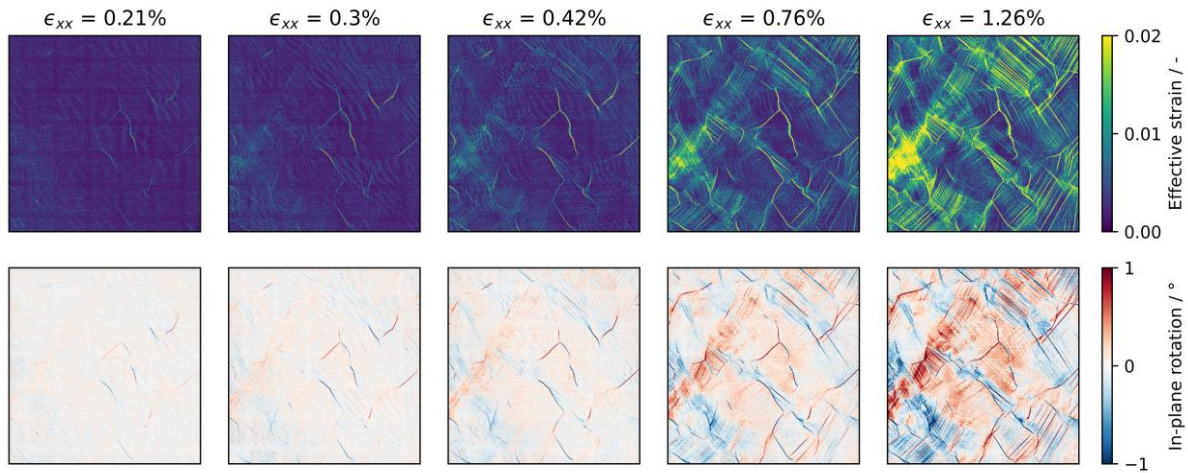

**Figure 3: Progression of grain boundary sliding for strain around the yield point. Five snapshots of strain (yellow points in (a)) are shown for strains immediately pre- and post-yield. The yield stress here is 120 MPa. A sub-region (b) of the ROI in Figure 2 is shown here for clarity. Effective strain and in-plane rotation maps clearly show the onset of grain boundary sliding prior to macroscopic yielding, followed by increasing levels of slip post-yield. Strain values indicated here are calculated using HRDIC, by taking the mean value of the *x*-component for each map.**

Unlike the combination of grain boundary sliding and opening observed by Mornout *et. al.* [15], no grain boundary opening is observed here either in the strain maps or in the raw images used for HRDIC. As such, there is no volume change or void formation associated with sliding at the strains examined here. Rotation maps in Figure 3c demonstrate how compatibility is maintained during grain boundary sliding; each boundary showing in-plane rotation in one direction is bounded by regions of the opposition direction. This is analogous to the behaviours observed previously in other materials with competing domains of slip (Zircaloy-4 [22], austenitic stainless steel [17]).

The qualitative observations in Figure 3 can also be quantified to determine the magnitudes of strains associated with different microstructural regions. By linking the HRDIC and EBSD datasets, we segment the time-series strain mapping data into grain core and grain boundary regions. To segment the map, grain boundary lines were dilated to include points within 1.5 µm of the boundary line. This value was selected to account for small misalignments between EBSD and HRDIC maps [20] and for the spatial smearing related to HRDIC measurements [33]. Grain boundaries can be further divided into Σ3 boundaries and other high angle grain boundaries (Figure 4a). Histograms of effective strain and in-plane rotation magnitude for four of the deformation snapshots in Figure 3 are plotted in Figure 4b.

The regions of most intense strain localisation are associated with grain boundaries for all deformation levels. This appears as both a greater probability of observing higher strain values (in the interval $0.01 < \epsilon_{eff} < 0.1$) in grain boundary regions and a shift in extreme values, with the 99.99$^{th}$ percentile strain values in grain boundaries far exceeding those in grain cores. The largest deviation between the two regions is detected prior to yield (Figure 4b, $\epsilon_{eff}$ = 0.3%), in agreement with the qualitative observations in Figure 3. With increasing levels of deformation, the strain probability distributions for all regions converge, showing an increase in the levels of slip within grain cores following yield.

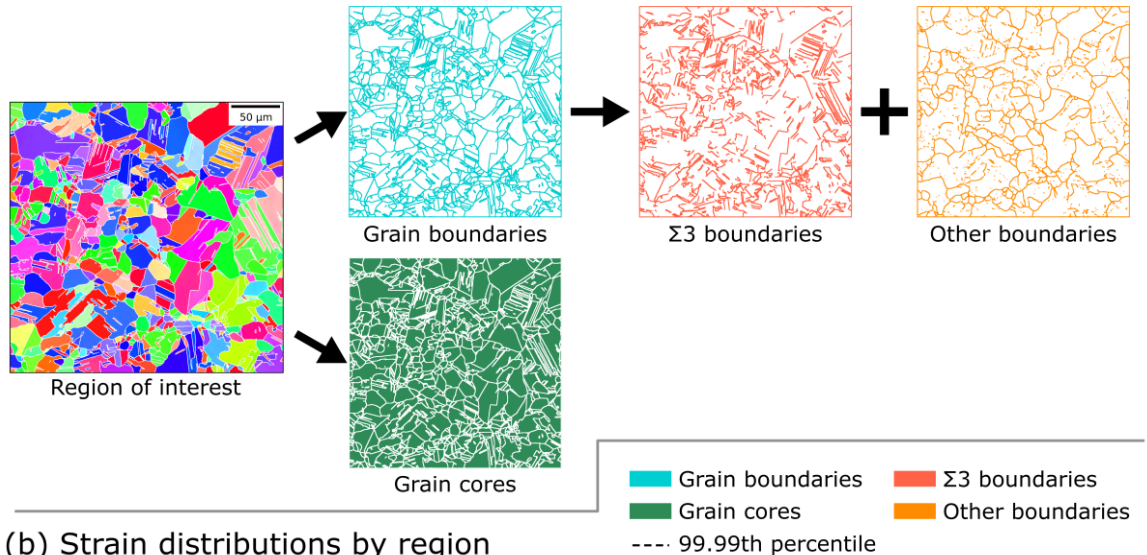

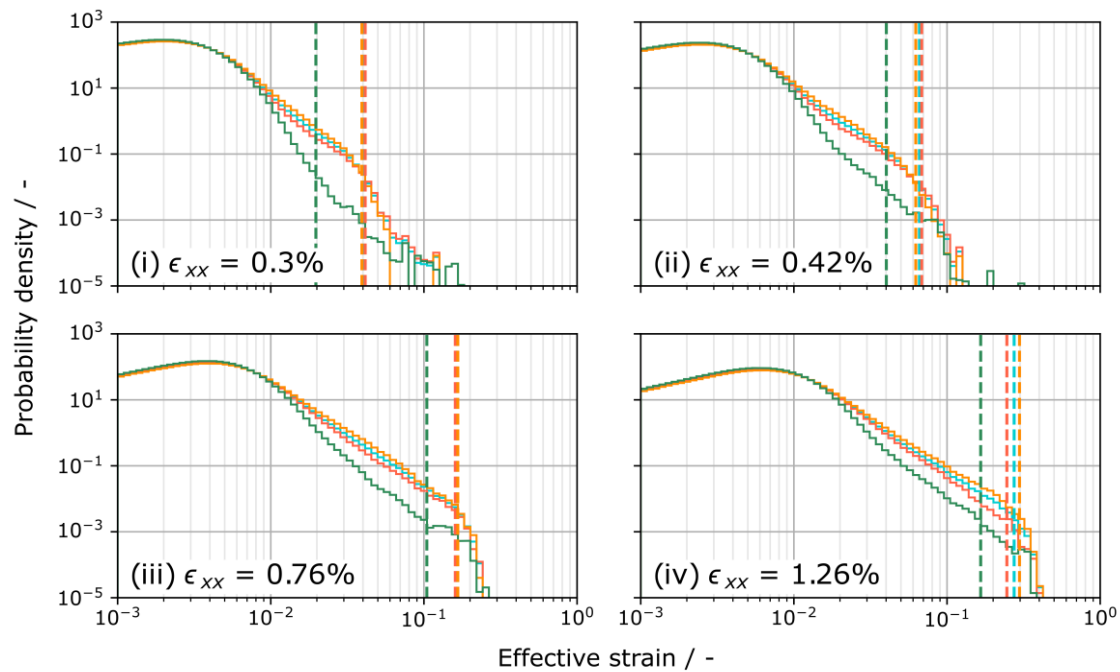

**Figure 4: Strain distributions partitioned by microstructural region. Histograms (b) show the progression of effective strain for four deformation steps around the yield point. Vertical dashed lines indicate the 99.99th percentile for each region.**

Further segmentation of grain boundary type into Σ3 (twin) and other high angle boundaries shows minimal quantitative difference in deformation intensity. Whilst Σ3 boundaries are associated with lower levels of strain for all deformation levels, the quantitative difference is small. Physically, it may be expected that the inherent disorder of non-coherent high angle grain boundaries would promote sliding, whilst the increased energetic penalty of sliding along a coherent boundary such as a coincident lattice site boundary would suppress sliding. The measurements presented here only weakly support this hypothesis.

Both slip and grain boundary sliding are inherently three-dimensional processes. Slip is crystallographically constrained to known deformation planes and directions and can be fully characterised from in-plane measurements using methods such as advanced slip trace analysis (i.e. the relative displacement ratio method [34]) or kinematical analyses (i.e. SSLIP [35,36]). Conversely, grain boundary sliding is only related to the morphology of the grain boundary. Recent studies have shown the value in height mapping using optical methods as a complementary technique to HRDIC [15,37–41]. Since grain boundary sliding has been found to have a significant out-of-plane component, linking these two techniques is necessary to fully quantify grain boundary sliding character [15,38].

Post-deformation height mapping was performed using a ZEISS LSM900 laser scanning confocal microscope (LSCM, C-Epiplan-Apochromat 50×/0.95 NA objective) with a lateral resolution of 62 nm. A median filter with a 15-pixel kernel size was applied to remove any height mapping artefacts related to the presence of the HRDIC speckle pattern. A least-squares planar surface was fitted to and removed from the measured surface to level the surface. A pre-deformation height map was not recorded as preliminary work found the surfaces prepared for EBSD had no-measurable height deviations from a flat plane as measured with this instrument. Due to differences in lateral resolution between LSCM and HRDIC, the LSCM dataset was resampled to the resolution of the strain map. A similar method to that used to align the EBSD and HRDIC datasets was then used to register the height map to the pre-deformation HRDIC reference frame [31], using nine manually selected homologous points and a projective transform. To manage small discrepancies in local registration between the maps, the map stack was binned by a factor of ten, taking the maximum value within each bin. Grains boundary regions were segmented from the stacked strain-height dataset using the aligned EBSD dataset.

The height gradient magnitude was found to be the most insightful quantity in examining out-of-plane deformation. Whilst this is not strictly a measure of strain, it is useful in identifying distinct changes in surface height that are typically associated with grain boundary sliding.

Figure 4: Strain distributions partitioned by microstructural region. Histograms (b) show the progression of effective strain for four deformation steps around the yield point. Vertical dashed lines indicate the 99.99th percentile for each region.Figure 4 shows the importance of considering extreme values to distinguish strains associated with sliding from those linked to slip. Accordingly, we define out-of-plane sliding as active in segments where the magnitude of the height gradient is above the $75^{th}$ percentile of all values in grain boundary regions. A similar criterion, also set at the $75^{th}$ percentile, is applied to the effective strain to identify in-plane sliding. We define regions not satisfying either criterion as showing minimal sliding activity. The results of this classification are detailed in Figure 5.

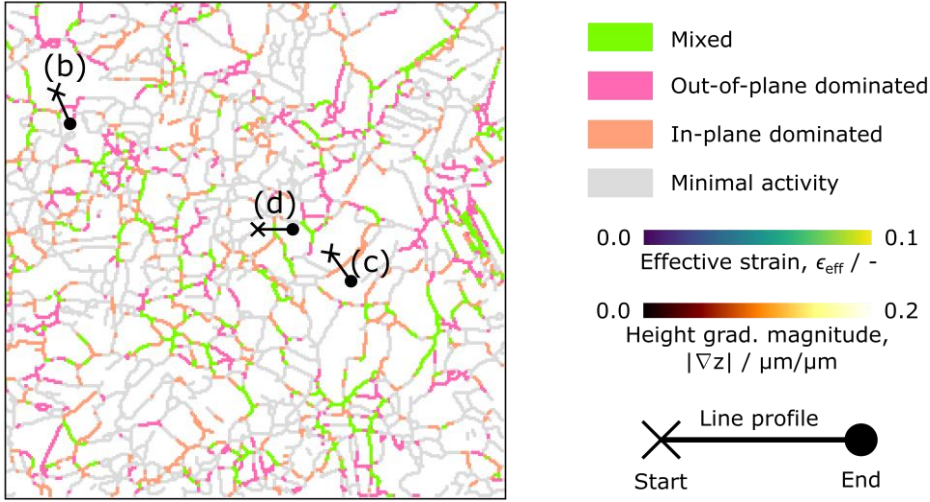
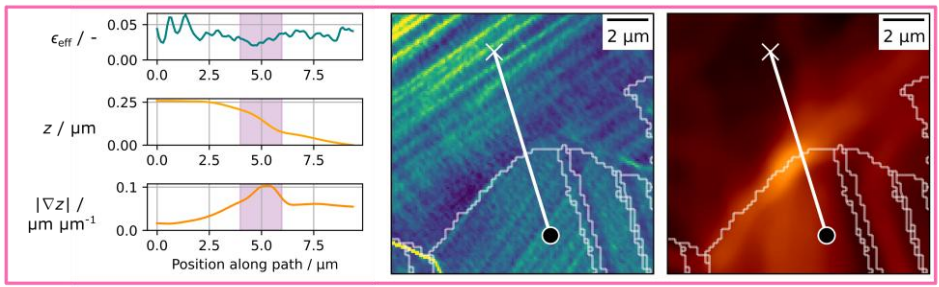
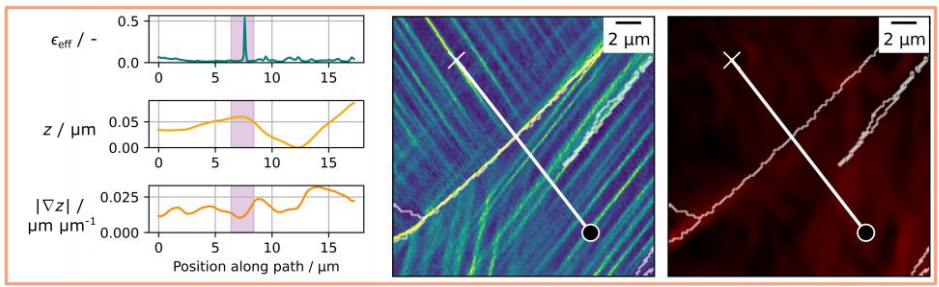
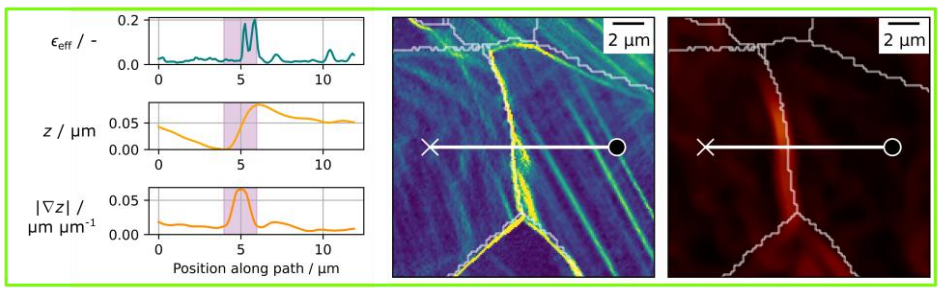

**Figure 5: Characterisation of grain boundary activity using combined strain and height mapping.** The GB activity map in (a) shows the deformation type at each grain boundary segment, including boundaries where no significant deformation was observed. Three separate line profiles have been extracted as examples of the three main grain boundary deformation modes of, (b) out-of-plane dominated sliding, (c) in-plane dominated sliding, and (d) mixed sliding.

As observed during early deformation, grain boundary sliding is widespread but does not occur on every boundary within the ROI. The activity map in Figure 5a shows that, whilst not every boundary slides, a significant minority of boundaries do have some degree of sliding character. The full spectrum of sliding character is observed across the ROI, from pure out-of-plane sliding (example in Figure 5b) to pure in-plane sliding (Figure 5c), as well as mixed character showing both in- and out-of-plane sliding occurring simultaneously (Figure 5d).

The prevalence of grain boundary sliding with pure out-of-plane character adds weight to the need to quantify this phenomenon with both HRDIC and optical profilometry. The example in Figure 5c shows one of many regions where there is no observable strain localisation found with HRDIC strain mapping, but there is significant out-of-plane motion detected with optical profilometry. As such, HRDIC is likely to underestimate the extent of grain boundary sliding.

There may be an expectation that grain boundary sliding is only observed as an artefact of the free surface. However, we argue that this is not the case for the following reason. If grain boundary sliding was primarily driven by the surface stress state, it would be expected that most sliding boundaries would have significant out-of-plane components as the out-of-plane direction represents that with the minimum level of constraint. However, the findings here do not support this hypothesis, as we observe a significant number of grain boundaries that have minimal out-of-plane motion associated with them, even with the lack of surface constraint. As such, we believe that some level of sliding would be expected within the bulk.

To summarise, we examine grain boundary sliding oxygen-free copper at room temperature using a combination of HRDIC strain mapping, *in-situ* testing within the SEM, and laser confocal microscopy height mapping. We find that grain boundary sliding is prevalent before macroscopic yield, prior to the onset of significant levels of deformation within cores. We quantify these trends, showing that the extreme strain values are associated with grain boundaries immediately pre- and post-yield. These observations are only possible due to the use of *in*-situ testing performed at a high temporal resolution. Finally, we determine the character of sliding, finding differing degrees of in-plane and out-of-plane sliding. These findings have significant implications for our understanding of copper deformation, in particular, for those who wish to build physics-based models of this materials system. These findings are particularly key for material failure modes such as fatigue and creep, where precursors to damage occur during the nominally elastic portion of deformation, where components typically operate.

**Acknowledgements**

The authors wish to thank the Applied Materials Technology group, the Design by Fundamentals team, and the Mechanics of Microstructures group at the University of Manchester for useful discussions and informal input throughout this project. All experimental work was performed in the Applied Materials Technology laboratory at the UK Atomic Energy Authority.

This work has been funded by STEP, a major technology and infrastructure programme led by UK Fusion Energy Ltd (UKFE), which aims to deliver the UK's prototype fusion powerplant and a path to the commercial viability of fusion. To obtain further information on the data and models underlying this paper please contact PublicationsManager@ukaea.uk*.

**CRediT statement**

**Ben Poole:** Conceptualisation, Methodology, Formal analysis, Investigation, Writing – Original Draft. **Dave Lunt**: Investigation, Writing – Review & Editing. **Luke Hewitt**: Investigation, Writing – Review & Editing. **Chris Hardie**: Investigation, Writing – Review & Editing, Supervision, Funding acquisition.

# References

[1] K. Pettersson, A study of grain boundary sliding in copper with and without an addition of phosphorus, J. Nucl. Mater. 405 (2010) 131–137. https://doi.org/10.1016/j.jnucmat.2010.07.044.

[2] T.G. Langdon, Grain boundary sliding revisited: Developments in sliding over four decades, J. Mater. Sci. 41 (2006) 597–609. https://doi.org/10.1007/s10853-006-6476-0.

[3] D. Texier, J. Milanese, M. Jullien, J. Genée, J.-C. Passieux, D. Bardel, E. Andrieu, M. Legros, J.-C. Stinville, Strain localization in the Alloy 718 Ni-based superalloy: From room temperature to 650 °C, Acta Mater. 268 (2024) 119759. https://doi.org/10.1016/j.actamat.2024.119759.

[4] M. Jullien, R.L. Black, J.C. Stinville, M. Legros, D. Texier, Grain size effect on strain localization, slip-grain boundary interaction and damage in the Alloy 718 Ni-based superalloy at 650 °C, Mater. Sci. Eng. A 912 (2024) 146927. https://doi.org/10.1016/j.msea.2024.146927.

[5] R.N. Stevens, Grain-Boundary Sliding in Metals, Metall. Rev. 11 (1966) 129–142. https://doi.org/10.1179/mtlr.1966.11.1.129.

[6] V. Tvergaard, Influence of grain boundary sliding on material failure in the tertiary creep range, Int. J. Solids Struct. 21 (1985) 279–293. https://doi.org/10.1016/0020-7683(85)90024-1.

[7] B. Poole, A. Marsh, D. Lunt, C. Hardie, M. Gorley, C. Hamelin, A. Harte, Nanoscale speckle patterning for combined high-resolution strain and orientation mapping of environmentally sensitive materials, Strain n/a (2024) e12477. https://doi.org/10.1111/str.12477.

[8] B. Poole, A. Marsh, D. Lunt, C. Hardie, M. Gorley, C. Hamelin, A. Harte, High-resolution strain mapping in a thermionic LaB6 scanning electron microscope, Strain n/a (2024). https://doi.org/10.1111/str.12472.

[9] B. Yavuzyegit, E. Avcu, A.D. Smith, J.M. Donoghue, D. Lunt, J.D. Robson, T.L. Burnett, J.Q. da Fonseca, P.J. Withers, Mapping plastic deformation mechanisms in AZ31 magnesium alloy at the nanoscale, Acta Mater. 250 (2023) 118876. https://doi.org/10.1016/j.actamat.2023.118876.

[10] A. Orozco-Caballero, D. Lunt, J.D. Robson, J. Quinta da Fonseca, How magnesium accommodates local deformation incompatibility: A high-resolution digital image correlation study, Acta Mater. 133 (2017) 367–379. https://doi.org/10.1016/j.actamat.2017.05.040.

[11] T. Dessolier, P. Lhuissier, F. Roussel-Dherbey, F. Charlot, C. Josserond, J.-J. Blandin, G. Martin, Effect of temperature on deformation mechanisms of AZ31 Mg-alloy under tensile loading, Mater. Sci. Eng. A 775 (2020) 138957. https://doi.org/10.1016/j.msea.2020.138957.

[12] M.A. Linne, T.R. Bieler, S. Daly, The effect of microstructure on the relationship between grain boundary sliding and slip transmission in high purity aluminum, Int. J. Plast. 135 (2020) 102818. https://doi.org/10.1016/j.ijplas.2020.102818.

[13] M.A. Linne, S. Daly, Data clustering for the high-resolution alignment of microstructure and strain fields, Mater. Charact. 158 (2019) 109984. https://doi.org/10.1016/j.matchar.2019.109984.

[14] A. Venkataraman, M. Linne, S. Daly, M.D. Sangid, Criteria for the prevalence of grain boundary sliding as a deformation mechanism, Materialia 8 (2019) 100499. https://doi.org/10.1016/j.mtla.2019.100499.

[15] C.J.A. Mornout, G. Slokker, T. Vermeij, D. König, J.P.M. Hoefnagels, SLIDE: Automated identification and quantification of grain boundary sliding and opening in 3D, Scr. Mater. 268 (2025) 116861. https://doi.org/10.1016/j.scriptamat.2025.116861.

[16] F. Briffod, T.E.J. Edwards, J.Q. Da Fonseca, J.-C. Stinville, D. Texier, T. Vermeij, Understanding strain localization in metallic materials: a review of high-resolution digital image correlation and related techniques, Sci. Technol. Adv. Mater. (2026) 2630488. https://doi.org/10.1080/14686996.2026.2630488.

[17] F. Di Gioacchino, J. Quinta da Fonseca, An experimental study of the polycrystalline plasticity of austenitic stainless steel, Int. J. Plast. 74 (2015) 92–109. https://doi.org/10.1016/j.ijplas.2015.05.012.

[18] T. Vermeij, C.J.A. Mornout, V. Rezazadeh, J.P.M. Hoefnagels, Martensite plasticity and damage competition in dual-phase steel: A micromechanical experimental–numerical study, Acta Mater. 254 (2023) 119020. https://doi.org/10.1016/j.actamat.2023.119020.

[19] A. Harte, M. Atkinson, M. Preuss, J. Quinta da Fonseca, A statistical study of the relationship between plastic strain and lattice misorientation on the surface of a deformed Ni-based superalloy, Acta Mater. 195 (2020) 555–570. https://doi.org/10.1016/j.actamat.2020.05.029.

[20] M.D. Atkinson, J.M. Donoghue, J.Q. da Fonseca, Measurement of local plastic strain during uniaxial reversed loading of nickel alloy 625, Mater. Charact. 168 (2020) 110561. https://doi.org/10.1016/j.matchar.2020.110561.

[21] D. Lunt, A. Orozco-Caballero, R. Thomas, P. Honniball, P. Frankel, M. Preuss, J. Quinta da Fonseca, Enabling high resolution strain mapping in zirconium alloys, Mater. Charact. 139 (2018) 355–363. https://doi.org/10.1016/j.matchar.2018.03.014.

[22] R. Thomas, D. Lunt, M.D. Atkinson, J. Quinta da Fonseca, M. Preuss, F. Barton, J. O'Hanlon, P. Frankel, Characterisation of irradiation enhanced strain localisation in a zirconium alloy, Materialia 5 (2019) 100248. https://doi.org/10.1016/j.mtla.2019.100248.

[23] Z. Ye, C. Li, M. Zheng, X. Zhang, X. Yang, J. Gu, In situ EBSD/DIC-based investigation of deformation and fracture mechanism in FCC- and L12-structured FeCoNiV high-entropy alloys, Int. J. Plast. 152 (2022) 103247. https://doi.org/10.1016/j.ijplas.2022.103247.

[24] D. Lunt, R. Thomas, M.D. Atkinson, A. Smith, R. Sandala, J.Q. da Fonseca, M. Preuss, Understanding the role of local texture variation on slip activity in a two-phase titanium alloy, Acta Mater. 216 (2021) 117111. https://doi.org/10.1016/j.actamat.2021.117111.

[25] M.P. Echlin, J.C. Stinville, V.M. Miller, W.C. Lenthe, T.M. Pollock, Incipient slip and long range plastic strain localization in microtextured Ti-6Al-4V titanium, Acta Mater. 114 (2016) 164–175. https://doi.org/10.1016/j.actamat.2016.04.057.

[26] M.E. Harr, S. Daly, A.L. Pilchak, The effect of temperature on slip in microtextured Ti-6Al-2Sn-4Zr-2Mo under dwell fatigue, Int. J. Fatigue 147 (2021) 106173. https://doi.org/10.1016/j.ijfatigue.2021.106173.

[27] A new approach to SEM in-situ thermomechanical experiments through automation, Ultramicroscopy 280 (2026) 114244. https://doi.org/10.1016/j.ultramic.2025.114244.

[28] A. Venkataraman, M.D. Sangid, A crystal plasticity model with an atomistically informed description of grain boundary sliding for improved predictions of deformation fields, Comput. Mater. Sci. 197 (2021) 110589. https://doi.org/10.1016/j.commatsci.2021.110589.

[29] D. Zukić, M. Jackson, D. Dimiduk, S. Donegan, M. Groeber, M. McCormick, ITKMontage: A Software Module for Image Stitching, Integrating Mater. Manuf. Innov. 10 (2021) 115–124. https://doi.org/10.1007/s40192-021-00202-x.

[30] J. Hirst, L. Sibson, A. Tayeb, B. Poole, M. Sampson, W. Bielajewa, M. Atkinson, A. Marsh, R. Spencer, R. Hamill, C. Hamelin, A. Harte, L. Fletcher, PYVALE: A Fast, Scalable, Open-Source 2D Digital Image Correlation (DIC) Engine Capable of Handling Gigapixel Images, (2026). https://doi.org/10.48550/arXiv.2601.12941.

[31] MechMicroMan/DefDAP, (2026). https://github.com/MechMicroMan/DefDAP (accessed February 12, 2026).

[32] J.P. Goulmy, D. Depriester, F. Guittonneau, L. Barrallier, S. Jégou, Mechanical behavior of polycrystals: Coupled in situ DIC-EBSD analysis of pure copper under tensile test, Mater. Charact. 194 (2022) 112322. https://doi.org/10.1016/j.matchar.2022.112322.

[33] F. Bourdin, J.C. Stinville, M.P. Echlin, P.G. Callahan, W.C. Lenthe, C.J. Torbet, D. Texier, F. Bridier, J. Cormier, P. Villechaise, T.M. Pollock, V. Valle, Measurements of plastic localization by heaviside-digital image correlation, Acta Mater. 157 (2018) 307–325. https://doi.org/10.1016/j.actamat.2018.07.013.

[34] Z. Chen, S.H. Daly, Active Slip System Identification in Polycrystalline Metals by Digital Image Correlation (DIC), Exp. Mech. 57 (2017) 115–127. https://doi.org/10.1007/s11340-016-0217-3.

[35] T. Vermeij, R.H.J. Peerlings, M.G.D. Geers, J.P.M. Hoefnagels, Automated identification of slip system activity fields from digital image correlation data, Acta Mater. 243 (2023) 118502. https://doi.org/10.1016/j.actamat.2022.118502.

[36] T. Vermeij, G. Slokker, C.J.A. Mornout, D. König, J.P.M. Hoefnagels, +SSLIP: Automated Radon-Assisted and Rotation-Corrected Identification of Complex HCP Slip System Activity Fields from DIC Data, Strain 61 (2025) e70000. https://doi.org/10.1111/str.70000.

[37] A. Rouwane, D. Texier, J.-N. Périé, J.-E. Dufour, J.-C. Stinville, J.-C. Passieux, High resolution and large field of view imaging using a stitching procedure coupled with distortion corrections, Opt. Laser Technol. 177 (2024) 111165. https://doi.org/10.1016/j.optlastec.2024.111165.

[38] M. Jullien, R.L. Black, J.C. Stinville, M. Legros, D. Texier, Grain size effect on strain localization, slip-grain boundary interaction and damage in the Alloy 718 Ni-based superalloy at 650 °C, Mater. Sci. Eng. A 912 (2024) 146927. https://doi.org/10.1016/j.msea.2024.146927.

[39] J.H. Liu, N. Vanderesse, J.-C. Stinville, T.M. Pollock, P. Bocher, D. Texier, In-plane and out-of-plane deformation at the sub-grain scale in polycrystalline materials assessed by confocal microscopy, Acta Mater. 169 (2019) 260–274. https://doi.org/10.1016/j.actamat.2019.03.001.

[40] A. Rouwane, D. Texier, S. Hémery, J.-C. Passieux, Q. Sirvin, J. Genée, A. Proietti, J.C. Stinville, Strain localization in Ti and Ti-alloys using three-dimensional topographic imaging, in: Issu Proc. 15th World Conf. Titan., Edinburgh, United Kingdom, 2023: p. 6 p. https://doi.org/10.7490/f1000research.1119929.1.

[41] W. Yin, F. Briffod, H. Hu, T. Shiraiwa, M. Enoki, Three-dimensional configuration of crystal plasticity in stainless steel assessed by high resolution digital image correlation and confocal microscopy, Int. J. Plast. 170 (2023) 103762. https://doi.org/10.1016/j.ijplas.2023.103762.